\documentclass{article}
\usepackage[a4paper]{geometry}
\usepackage[T1]{fontenc}

\usepackage[]{amsmath, amssymb, amsthm, tabularx}

\usepackage[]{a4, xcolor, here}

\usepackage[]{pstricks, pst-text, pst-node, pst-tree, gastex}

\usepackage{tikz}

\usepackage[tikz]{bclogo}

\usepackage{appendix}

\usepackage{cite}

\usepackage{boites,graphicx}

\usepackage{soul}

\definecolor{pastelyellow}{rgb}{0.99, 0.99, 0.59}
\definecolor{aqua}{rgb}{0.0, 1.0, 1.0} 
\definecolor{aquamarine}{rgb}{0.5, 1.0, 0.83} 
\definecolor{bananayellow}{rgb}{1.0, 0.88, 0.21}
\definecolor{burgundy}{rgb}{0.5, 0.0, 0.13}
\definecolor{ao(english)}{rgb}{0.0, 0.5, 0.0}

\setul{}{0.2ex}
\setulcolor{bananayellow}

\usepackage[normalem]{ulem}

\usepackage{mathrsfs}

\usepackage{stmaryrd}

\usepackage{enumerate}

\usepackage{paralist}

\usepackage{multienum}

\usepackage{multicol}

\usepackage{fancyhdr}

\usepackage[]{hyperref} 
\hypersetup{
		colorlinks = true,
		linkcolor = red,
		anchorcolor = black,
		citecolor = blue, % green,
		filecolor = cyan,
		menucolor = red,
		runcolor = cyan,
		urlcolor = blue,
		linkbordercolor = {white},
		linktocpage = true
}

\newtheorem{theorem}{Theorem}[section]
\newtheorem{proposition}[theorem]{Proposition}
\newtheorem{lemma}[theorem]{Lemma}
\newtheorem{corollary}[theorem]{Corollary}

\theoremstyle{definition}
\newtheorem{definition}[theorem]{Definition}

\newtheorem{example}[theorem]{Example}

\newtheorem{remark}[theorem]{Remark}

\makeatletter
\def\thmhead@plain#1#2#3{%
  \thmname{#1}\thmnumber{\@ifnotempty{#1}{ }\@upn{#2}}%
  \thmnote{ {\the\thm@notefont#3}}}
\let\thmhead\thmhead@plain
\makeatother

\newcommand{\bC}{\mathbf{C}}

\newcommand{\cC}{\mathcal{C}}

\newcommand{\cF}{\mathcal{F}}
\newcommand{\cG}{\mathcal{G}}

\newcommand{\cU}{\mathcal{U}}
\newcommand{\cV}{\mathcal{V}}

\newcommand{\cX}{\mathcal{X}}
\newcommand{\cY}{\mathcal{Y}}

\newcommand{\GL}{\mathrm{GL}}
\newcommand{\GaL}{\Gamma\mathrm{L}}

\newcommand{\Aut}{\mathrm{Aut}}
\newcommand{\SAut}{\mathrm{SAut}}
\newcommand{\rsp}{\mathrm{rowsp}}
\newcommand{\bbF}{{\mathbb F}} 

\renewcommand{\geq}{\geqslant}
\renewcommand{\leq}{\leqslant}

\newcommand{\then}{\Longrightarrow}

\newcommand{\en}{\longrightarrow}

{\end{list}}%

\begin{document}

\renewcommand{\headrulewidth}{0pt}

\rhead{ }
\chead{\scriptsize Equivalence for Flag Codes}
\lhead{ }

\title{Equivalence for Flag Codes
\renewcommand\thefootnote{\arabic{footnote}}\footnotemark[1] }

\author{\renewcommand\thefootnote{\arabic{footnote}}
Miguel \'Angel Navarro-P\'erez\footnotemark[2], 
\renewcommand\thefootnote{\arabic{footnote}} 
 Xaro Soler-Escriv\`a\footnotemark[3]}

\footnotetext[1]{The authors received financial support of Ministerio de Ciencia e Innovaci\'on (PID2022-142159OB-I00) and Conselleria de Innovaci\'on, Universidades, Ciencia y Sociedad Digital (CIAICO/2022/167).}

\footnotetext[2]{Dpt.\ de Matem\'aticas, Universidad Carlos III de Madrid, Av. de la Universidad, 30, 28911 Legan\'es, Madrid (Spain).}

\footnotetext[3]{Dpt.\ de Matem\`atiques, Universitat d'Alacant, Sant Vicent del Raspeig, Ap.\ Correus 99, 03080 Alacant (Spain). \\\  E-mail adresses: \texttt{mignavar@math.uc3m.es, xaro.soler@ua.es.}}

{\small \date{\today}} 

\maketitle

%{
%\footnotesize 
%\tableofcontents
%}

\begin{abstract}
Given a finite field $\bbF_q$ and a positive integer $n$, a \emph{flag} is a sequence of nested $\bbF_q$-subspaces of a vector space $\bbF_q^n$ and a \emph{flag code} is a nonempty collection of flags.
The \emph{projected codes} of a flag code are the constant dimension codes containing all the subspaces of prescribed dimensions that form the flags in the flag code. 

In this paper we address the notion of equivalence for flag codes and explore in which situations such an equivalence can be reduced to the equivalence of the corresponding projected codes. 
In addition, this study leads to new results concerning the automorphism group of certain families of flag codes, some of them also introduced in this paper.
\end{abstract}



%%%%%%%%%%%%%%%%%%%%
%                  %
%   Introduction   %
%                  %
%%%%%%%%%%%%%%%%%%%%

\section{Introduction}\label{sec:Introduction}

\emph{Network coding} was introduced in \cite{AhCaiLiYeung2000} as a new method to send information using a network modeled as an acyclic directed multigraph with several senders and
receivers, where intermediate nodes are allowed to perform linear combinations of the received packets, instead of simply routing them. The first algebraic approach to network coding was presented in \cite{KoetKschi08}. In that paper, the authors propose using vector spaces over a finite field $\bbF_q$ as codewords. In this context, \emph{subspace codes} arise as sets of subspaces of a prescribed vector space $\bbF_q^n$. In particular, if every codeword, i.e., every subspace, has the same dimension, we speak about \emph{constant dimension codes} of $\bbF_q^n.$ During the last decade, there have been many different works on this topic. We refer the reader to \cite{TrautRos2018} and the references therein for an overview on constant dimension codes.

\emph{Flag codes} in the network coding setting were introduced by Liebhold, Nebe and Vazquez-Castro in \cite{LiebNebeVaz2018} as a generalization of constant dimension codes. A \emph{flag} on $\bbF_q^n$ is a sequence of nested subspaces of $\bbF_q^n$. The list of dimensions of the subspaces in a flag is called its \emph{type vector} and a \emph{flag code} $\cC$ is a nonempty collection of flags (of the same type) on $\bbF_q^n$. 
Constant dimension codes and flag codes are closely related by the notion of \emph{projected codes}. For each dimension in the type vector of a flag code $\cC$,  we consider the constant dimension code composed by all the subspaces of this dimension appearing in some flag in $\cC$. This list of constant dimension codes is the set of {\em projected codes} associated to the flag code $\cC$ (see \cite{CasoPlanar}). 
During the last years, some problems on flag codes have been tackled using this close relationship. For instance, papers \cite{Cl-MA2020, PlanarOrbConstr, CasoPlanar, CasoGeneral, Singers},  provide several characterizations of flag codes with maximum distance (\emph{optimum distance flag codes}) in terms of their projected codes. These characterizations have allowed to provide systematic constructions of them by using different tools. In \cite{Consistent}, the authors study under which conditions the parameters, as well as other structural properties, of a flag code can be obtained from the ones of their projected codes. On the other hand, in \cite{Cotas} the reader can find bounds for the cardinality of flag codes with prescribed minimum distance. The techniques used to obtain these bounds also involve the projected codes. Other works such as \cite{AlfNeriZullo2023, FouNebe2023, Kurz20} study flag codes from other points of view, regardless of their projected codes.

A general problem in Coding Theory is determining when two different codes are equivalent, that is, when they are essentially the same. In \cite{Traut13} this question is addressed for the first time for constant dimension codes and the notions of (semi)linearly equivalent codes and the automorphism group of a constant dimension code arise. After that paper, different authors have studied this problem for specific families of constant dimension codes in \cite{CasPolSanZullo2023, GLLehmann2023, Zullo2023}. Nevertheless, in the context of flag codes, determining whether two flag codes are equivalent is an unexplored question yet. In the present paper, we bring the concepts of linear and semilinear equivalence to the flag codes scenario. As in the papers mentioned in the previous paragraph, here we also make use of the powerful relationship between a flag code and its projected codes. However, in this case the original point of view is reversed. We notice that different flag codes can share exactly the same list of projected codes. This is the reason why the property of being equivalent cannot been easily translated from the list of projected codes to the initial flag code. We observe that this problem can be solved by changing the perspective. We start characterizing those families of constant dimension codes that can provide flag codes and then, we identify when a set of constant dimension codes generates a unique flag code. This innovative approach allows us to answer some questions related to flag code equivalence and also to give a sufficient condition to compute the automorphism group of a flag code as the intersection of the ones of its projected codes, which is not true for general flag codes.

The paper is organized as follows. In Section \ref{sec: preliminaries} we give the basic background on both constant dimension codes and flag codes, putting special attention to the notions of linear and semilinear equivalent constant dimension codes and the one of projected codes of a flag code. In Section \ref{sec: fc generated by sets of cdc}, we study under which conditions a set of constant dimension codes is suitable to produce flag codes. Moreover, we define and analyze a new property on flag codes: the  \emph{subspace-inclusion-closed} property (SIC, for short) and use it to characterize those families of constant dimension codes that generate one, and only one, flag code. In Section \ref{sec: l and sl eq}, we introduce the notions of (semi)linear equivalence for flag codes and give sufficient conditions to reduce the flag code equivalence problem to the corresponding problem for their projected codes. This study has consequences for the computation of the (semi)linear automorphism group of a flag code that, under certain conditions, turns to be the intersection of the ones of the projected codes. Last, Section \ref{sec: examples} is devoted to illustrate the theoretical results obtained in the preceding sections. For this purpose, we present two families of flag codes -the ones of \emph{increasing/decreasing} flag codes- that, in many cases, enclose the one of optimum distance flag codes. We prove increasing/decreasing flag codes are equivalent if, and only if, their projected codes are equivalent. Moreover, computing their automorphism group just requires to know the ones of their projected codes. 

%%%%%%%%%%%%%%%%%%%%%%%%%%%%%%%%%%%%%%%%%%%%%

\section{Preliminaries}\label{sec: preliminaries}

In this section, we recall the basic background on constant dimension codes and flag codes that we will need throughout the paper. First, we recall the definition of constant dimension code and focus on some concepts, such as the subspace distance and linear and semilinear equivalence, related to this kind of codes. Later, we will move to the context of flag codes and we will bring back the notions of flag distance and projected codes, which will  play an important role in subsequent sections.

%%%%%%%%%%%%%%%%%%%%%%%%%%%%%%%%%%%%%%%%%%%%%

\subsection{Constant dimension codes}\label{subsec:constdimcod}

Given a finite field $\bbF_q$ and integers $1\leq k < n$, the set of all $k$-dimensional vector subspaces of $\bbF_q^n$  over $\bbF_q$ is called the {\em Grassmann variety} and will be denoted by $\cG_q(k, n)$. We put $\bbF_q^{k\times n}$ for the set of all $(k\times n)$-matrices with entries in $\bbF_q$ and $\GL_n(\bbF_q)$ for the {\em general linear group} of degree $n$ over $\bbF_q$, which is the set of all invertible matrices in $\bbF_q^{n\times n}$. Every subspace $\cU\in \cG_q(k, n)$ is generated by the rows of a full-rank matrix $U\in \bbF_q^{k\times n}$, that is, $\cU=\rsp(U)$. 

The Grassmann variety can be endowed of a metric by considering the \emph{subspace distance}:
\begin{equation}\label{eq:dist}
d_S(\cU,\cV)=\dim(\cU+\cV)-\dim(\cU\cap \cV)=2(k -\dim(\cU\cap \cV)),
\end{equation}
for all $\cU, \cV\in\cG_q(k, n)$  (see \cite{KoetKschi08}). Using this metric, a nonempty subset $\bC$ of $\cG_q(k, n)$ is called a \emph{constant dimension code}. The \emph{minimum distance} of $\bC$ is
\begin{equation}\label{eq:distC}
d_S(\bC)= \min\{ d_S(\cU, \cV) \ | \ \cU, \cV\in \bC, \ \cU\neq \cV\}.
\end{equation}
In case that $|\bC|=1$, we put $d_S(\bC)=0$. According to (\ref{eq:dist}), the value $d_S(\bC)$ is always an even integer satisfying
\begin{equation}\label{eq: bound distC}
0\leq d_S(\bC)
\leq 
\min\{2k, 2(n-k)\}
=
\left\{\begin{array}{lll}
2k  & \mbox{if} & 2k\leq n,\\
2(n-k) & \mbox{if} & 2k\geq n.
\end{array}
\right.
\end{equation}
Codes attaining the previous upper bound are called {\em constant dimension codes of  maximum distance}. The next result states when two $k$-dimensional subspaces of $\bbF_q^n$ give the maximum possible distance. The proof is straightforward and we omit it.
\begin{lemma}\label{lem: max dist}
Let $\cU$ and $\cV$ be subspaces of $\bbF_q^n$ of dimension $1\leq k< n$. The following statements hold:
\begin{enumerate}
\item If $1\leq k\leq \frac{n}{2},$ then $d_S(\cU, \cV) = 2k$ if, and only if, $\cU\cap\cV=\{0\}$.
\item If $\frac{n}{2}\leq k < n,$ then $d_S(\cU, \cV) = 2(n-k)$ if, and only if, $\cU+\cV=\bbF_q^n$.
\end{enumerate} 
\end{lemma}

We are interested in the action of two specific groups on $\cG_q(k, n)$. On the one hand, for any $A\in \GL_n(\bbF_q)$ and $\cU=\rsp(U)\in \cG_q(k, n)$, the following map defines an action of the general linear group on the Grassmann variety (see \cite{TrautManRos2010}): 
\begin{equation}\label{eq:acGL}
\cU\cdot A=\rsp (UA)\in \cG_q(k, n).
\end{equation}
On the other hand, as an extension of  $\GL_n(\bbF_q)$, we consider the set all invertible semilinear maps of $\bbF_q^n$, that is, the {\em general semilinear group} $\GaL_n(\bbF_q)$. This group can be factorised as the semidirect product $\GaL_n(\bbF_q)=\GL_n(\bbF_q)\rtimes \Aut(\bbF_q)$, where $\Aut(\bbF_q)$ is  the automorphism group of $\bbF_q$. For any $A, B\in\GL_n(\bbF_q)$ and $\varphi, \varphi'\in \Aut(\bbF_q)$, the multiplication in $\GaL_n(\bbF_q)$, seen as the previous semidirect product, is: 
\[
(A,\varphi)(B,\varphi')=(A\varphi^{-1}(B), \varphi\varphi'),
\]
where $\varphi^{-1}$ applies to $B$ element-wise. Now, for any $(A,\varphi)\in \GaL_n(\bbF_q)$ and $\cU=\rsp(U)\in \cG_q(k, n)$, the map
\begin{equation}\label{eq:acGaL}
\cU\cdot (A,\varphi)=\rsp (\varphi(UA))
\end{equation}
defines an action of $\GaL_n(\bbF_q)$ on $\cG_q(k, n)$ (see \cite{Traut13}).

Notice that, starting from a given constant dimension code $\bC\subseteq \cG_q(k, n)$, the actions  defined in (\ref{eq:acGL}) and (\ref{eq:acGaL}) of $\GL_n(\bbF_q)$ and $\GaL_n(\bbF_q)$ respectively on the Grassmann variety, allow to obtain new constant dimension codes, namely, 
\[
\bC\cdot A=\{\cU\cdot A\ |\  \cU\in \bC\} \quad \mbox{ and } \quad \bC\cdot (A,\varphi)=\{\cU\cdot (A,\varphi)\ |\  \cU\in \bC\},
\]
for any $A\in \GL_n(\bbF_q)$ and $\varphi \in \Aut(\bbF_q)$. Clearly, these two new codes have the same cardinality than $\bC$. Moreover, the action of both $ \GL_n(\bbF_q)$ and $\GaL_n(\bbF_q)$ on the Grassmann variety respects the subspace distance defined in (\ref{eq:dist}),  (see \cite{Traut13, TrautManRos2010}). This leads to the following definitions of code equivalence, which were extensively studied in \cite{Traut13}. 

\begin{definition}\label{def:const_equiv}
We say that two constant dimension codes $\bC, \bC'\subseteq \cG_q(k, n)$ are {\em linearly equivalent} (resp. {\em semilinearly equivalent}) if there exists $A\in \GL_n(\bbF_q)$ (resp. $(A,\varphi)\in \GaL_n(\bbF_q)$) such that $\bC'=\bC\cdot A$ (resp. $\bC'=\bC\cdot (A,\varphi)$).
\end{definition}

Linearly equivalent codes are, in particular semilinearly equivalent. Moreover, notice that for any $\lambda \in \bbF_q^\ast$, the scalar matrix $\lambda I_n$ fixes any vector subspace in $\bbF_q^n$. In particular, we obtain that $\lambda I_n$ fixes any constant dimension code $\bC$, that is $\bC\cdot (\lambda I_n)=\bC$. Thus, it can be said that scalar matrices are (linear) automorphisms of any code $\bC$. More generally, we have the following definition:

\begin{definition}\label{def:aut_constdim} (\cite[Def. 2.1]{Traut13})
Given a constant dimension code $\bC$, its {\em linear automorphism group} is defined as  $\Aut(\bC)=\{A\in  \GL_n(\bbF_q)\ |\ \bC\cdot A=\bC\}$ and its {\em semilinear automorphism group}  is $\SAut(\bC)=\{(A,\varphi)\in \GaL_n(\bbF_q)\ |\ \bC\cdot (A,\varphi)=\bC\}$.
\end{definition}

\begin{remark}
For any constant dimension code $\bC$, one can easily see that $\Aut(\bC)$ is a subgroup of $\SAut(\bC)$.
\end{remark}

%%%%%%%%%%%%%%%%%%%%%%%%%%%%%%%%%%%%%%%%%%%%%

\subsection{Flag codes}\label{sec:backgroundFlags}

Flag codes appear for the first time in \cite{LiebNebeVaz2018}, where the authors present them as a generalization of constant dimension codes in network coding. In this case, codewords are flags, that is, sequences of nested subspaces of some prescribed dimensions of a vector space $\bbF_q^n$. Since then, several papers on this subject have proliferated (see, for instance, \cite{AlfNeriZullo2023,Cl-MA2020,Consistent,PlanarOrbConstr,Cotas,CasoPlanar,CasoGeneral, FouNebe2023, Kurz20, Singers}).  We recall here the definitions and results that we need for the present paper. 

Throughout this work, we will consider positive integers $0<t_1< \dots < t_r < n$. A {\em flag} of type $(t_1, \dots, t_r)$ on $\bbF_q^n$ is a sequence 
$\mathcal{F}=(\mathcal{F}_1, \dots,  \mathcal{F}_r)$ where
\[
\{0\}\subsetneq \mathcal{F}_1 \subsetneq \cdots \subsetneq \mathcal{F}_r \subsetneq \mathbb{F}_q^n,
\]
and $\mathcal{F}_i$ is a vector subspace of $\bbF_q^n$ of dimension $t_i$, for all $1\leq i\leq r$. 

The set of flags of type $(t_1, \dots, t_r)$ on $\mathbb{F}_q^n$ is known as the {\em flag variety} of type $(t_1, \dots, t_r)$ on $\mathbb{F}_q^n$  and we will denote it by $\mathcal{F}_q((t_1,\dots, t_r),n)$. This flag variety can be endowed with a metric by extending the subspace distance defined in (\ref{eq:dist}). Given two flags  $\cF=(\mathcal{F}_1,\dots,  \mathcal{F}_r)$ and $\cF'=(\mathcal{F}'_1, \dots, \mathcal{F}'_r)$ in $\mathcal{F}_q( (t_1, \dots, t_r),n)$, their {\em flag distance} is defined as
\begin{equation}\label{eq: flag dist}
d_f(\cF,\cF')= \sum_{i=1}^r d_S(\mathcal{F}_i, \mathcal{F}'_i).
\end{equation}

A {\em flag code of type $(t_1,\dots,t_r)$} on $\bbF_q^n$ is a nonempty subset 
$$
\cC\subseteq \mathcal{F}_q((t_1,\dots, t_r),n) \subset \cG_q(t_1, n)\times \dots\cG_q(t_r, n)
$$
and its {\em minimum distance}  is given by
\[
d_f(\cC)=\min\{d_f(\cF,\cF')\ |\ \cF,\cF'\in\cC, \ \cF\neq \cF'\}.
\]
As for constant dimension codes, if $|\cC|=1$, we put $d_f(\cC)=0$. As a consequence of (\ref{eq: bound distC}) together with (\ref{eq: flag dist}), the minimum distance $d_f(\cC)$ is always a positive even integer satisfying
\begin{equation}\label{eq: bound flag dist}
0\leq d_f(\cC) \leq \sum_{i=1}^r \min\{2t_i, 2(n-t_i)\}.
\end{equation}
If $d_f(\cC)$ attains this upper bound we say that $\cC$ is an \emph{optimum distance flag code}. These codes have been characterized in \cite{CasoPlanar, Singers} in terms of the next family of constant dimension codes.  Given a flag code $\cC$ of type $(t_1, \dots, t_r)$ on $\bbF_q^n,$ for every $1\leq i\leq r$, the  constant dimension code 
\begin{equation}\label{eq:def_proj}
\cC_i=\{ {\cF}_i \ |\   (\mathcal{F}_1,\dots,  \mathcal{F}_r) \in \cC \}\subseteq \cG_q(t_i,n)
\end{equation}
is said to be the {\em $i$-th projected code} of $\cC$ (see \cite{CasoPlanar}).  It is clear that $|\cC_i|\leq |\cC|$. In the special case in which $|\cC_i|=|\cC|,$ for every $1\leq i\leq r$, we say the flag code $\cC$ is \emph{disjoint}. A more general notion of projected codes of a flag code and of disjointness can be found in \cite{Cotas}.

The following result characterizes optimum distance flag codes in terms of (some of) their projected codes. To this end, we consider two remarkable dimensions $t_a$ and $t_b$ in the type vector $(t_1, \dots, t_r)$, defined as

\begin{equation}\label{eq: dim ta and tb}
t_a=\max\{ t_i \ | \ 2t_i\leq n\} \ \ \text{and} \ \  t_b=\min\{ t_i \ | \ 2t_i\geq n\}.
\end{equation}

Notice that, if $2t_r\leq n$, then $t_a=t_r$ and $t_b$ does not exist. On the other hand, if $2t_1\geq n$, then $t_b=t_1$ and $t_a$ does not exist. In the rest of situations, dimensions $t_a$ and $t_b$ are well defined and, in general,  are different and consecutive dimensions in the type vector, except for those type vectors containing the dimension $\frac{n}{2}$. In such a situation, we have $t_a=t_b=\frac{n}{2}$.

\begin{theorem}(\cite[Th. 3.11]{CasoPlanar} and \cite[Th. 4.8]{Singers})\label{th: char odfc}
    Let $\cC$ be a flag code of type $(t_1, \dots, t_r)$ on $\bbF_q^n.$ The following statements are equivalent:
    \begin{enumerate}
        \item $\cC$ is an optimum distance flag code.
        \item $\cC$ is disjoint and  $\cC_i$ has maximum distance for every $1\leq i\leq r$. 
        \item $|\cC|=|\cC_i|$ and  $\cC_i$ has maximum distance for $i\in\{a, b\}$ defined  in (\ref{eq: dim ta and tb}). 
    \end{enumerate}
\end{theorem}

%%%%%%%%%%%%%%%%%%%%%%%%%%%%%%%%%%%%%%%%%%%%%

\section{Flag codes generated by sets of constant dimension codes}\label{sec: fc generated by sets of cdc}

In this section, we study under which conditions a family of constant dimension codes produces flag codes. In that situation, we realise that different flag codes can come from the same family of constant dimension codes. Then we give a method to obtain the largest flag code generated by a family of constant dimension codes and we determine in which cases such a family gives exactly one flag code. These notions will be crucial in Section \ref{sec: l and sl eq} to properly connect the (semi)linear equivalence of flag codes with the (semi)linear equivalence of their projected codes.

\begin{definition}
A  Cartesian product of constant dimension codes $\bC_1\times \dots \times \bC_r\subseteq   \cG_q(t_1,n)\times \dots \times \cG_q(t_r,n)$, is said to be a {\em generating set of flag codes} of type $(t_1,\dots, t_r)$ if there exists a flag code $\cC\subseteq \mathcal{F}_q((t_1,\dots, t_r),n)$ such that $\cC_i=\bC_i$ for all $1\leq i\leq r$. In this case, we say that the flag code $\cC$ is generated by the set $\bC_1\times \dots\times\bC_r$.
\end{definition}

Notice that not every Cartesian product of constant dimension codes $\bC_1\times \dots \times \bC_r\subseteq   \cG_q(t_1,n)\times \dots \times \cG_q(t_r,n)$ allows the generation of flag codes, as we can see in the next example.

\begin{example}
    Consider the standard $\bbF_q$-basis $\{e_1, e_2, e_3\}$ of $\bbF_q^3$ and the Cartesian product $\bC_1\times\bC_2\subseteq \cG_q(1, 3)\times\cG_q(2, 3)$ with 
    $$
    \bC_1=\{\langle e_1\rangle, \langle e_1+e_3\rangle\} \ \text{and} \ \bC_2=\{\langle e_1+e_2, e_3\rangle, \langle e_2, e_3\rangle\}.
    $$
    It is easy to see that no flag $(\cF_1, \cF_2)$ of type $(1,2)$ on $\bbF_q^3$ can be formed with $\cF_1\in\bC_1$ and $\cF_2\in\bC_2$.
    \end{example}

The above example illustrates that the following condition is necessary for a set of constant dimension codes to generate flag codes.

\begin{remark}\label{rem: generating set then nonempty}
Let $\bC_1\times \dots \times \bC_r\subseteq   \cG_q(t_1,n)\times \dots \times \cG_q(t_r,n)$  be a generating set of flag codes. Then 
$$
(\bC_1\times \dots \times \bC_r)\cap\cF_q((t_1, \dots, t_r), n) \neq \emptyset.
$$
\end{remark}

Nevertheless, this condition is not sufficient for a Cartesian product of constant dimension codes to be a generating set of flag codes, as the next example reflects.

\begin{example}
Let $\{e_1, e_2, e_3\}$ be the standard $\bbF_q$-basis of $\bbF_q^3$ and consider the Cartesian product $\bC_1\times\bC_2\subseteq \cG_q(1, 3)\times\cG_q(2, 3)$ with 
    $$
    \bC_1=\{\langle e_1\rangle, \langle e_2\rangle\} \ \text{and} \ \bC_2=\{ \langle e_2, e_1+e_3\rangle, \langle e_2, e_3\rangle\}.
    $$
    Then 
    $$
    (\bC_1\times\bC_2)\cap\cF_q((1,2),3) =\{\cF, \cF'\},
    $$
where $\cF=(\langle e_2\rangle, \langle e_2, e_1+e_3\rangle)$ and $\cF'=(\langle e_2\rangle, \langle e_2, e_3\rangle)$.    However, there is no flag code $\cC\subseteq \cF_q((1,2),3)$ with $\bC_1$ and $\bC_2$ as its projected codes, since $\langle e_1\rangle\in\bC_1$ is not included in any subspace in $\bC_2.$ Thus, $\bC_1\times\bC_2$ is not a generating set of flag codes.
\end{example}

The following result characterizes those Cartesian products of constant dimension codes that are generating sets of flag codes. 

\begin{theorem}\label{theo:carac_seqgen}
Let $\bC_1\times \dots \times \bC_r\subseteq   \cG_q(t_1,n)\times \dots \times \cG_q(t_r,n)$ be a Cartesian product of constant dimension codes. The following sentences are equivalent:
\begin{enumerate}[$(i)$]
\item $\bC_1\times \dots\times \bC_r$ is a generating set of flag codes on $\bbF_q^n$ of type $(t_1,\dots, t_r)$.

\item The set of flags  $\cC=(\bC_1\times \dots\times \bC_r)\cap \cF_q((t_1, \dots, t_r), n)$ is nonempty and it satisfies $\cC_i=\bC_i$ for all $1\leq i\leq r.$

\item For each $i\in \{1,\dots , r\}$ and each subspace $\cU_i\in\bC_i$, there exist subspaces $\cU_{i-1}\in\bC_{i-1}$ and $\cU_{i+1}\in\bC_{i+1}$ such that $\cU_{i-1}\subsetneq \cU_i\subsetneq \cU_{i+1}
,$ (considering $\bC_0=\{\{0\}\}$ and $\bC_{r+1}=\{\bbF_q^n\}$).

\item For each $i\in \{1,\dots , r\}$ and each subspace $\cU_i\in\bC_i$, there exists a flag $\cF\in\mathcal{F}_q((t_1,\dots, t_r),n)$ such that $\cF_i=\cU_i$ and $\cF_j\in\bC_j$, for all $1\leq j\leq r$.
\end{enumerate}
\end{theorem}
\begin{proof}
$ $\newline

$\boxed{(i)\then (ii)}$ By definition of generating set of flag codes, there exists a flag code $\emptyset \neq \cC'\subseteq\cF_q((t_1, \dots, t_r), n)$ such that $\cC'_i=\bC_i,$ for every $1\leq i\leq r.$ Notice that
$$
\emptyset\neq\cC'\subseteq\cC=(\bC_1\times \dots\times \bC_r)\cap \cF_q((t_1, \dots, t_r), n).
$$ 
Then, for every $1\leq i\leq r,$ we have $\bC_i=\cC'_i\subseteq\cC_i\subseteq\bC_i$, i.e., $\cC_i=\bC_i.$

$\boxed{(ii) \then (iii)}$
Now, given $1\leq i\leq r$, since $\bC_i=\cC_i$, for any subspace $\cU_i\in \bC_i$, we can consider a flag $\cF\in\cC$ such that $\cF_i=\cU_i$. Then, putting $\cF_0=\{0\}$ and $\cF_{r+1}=\bbF_q^n$, we always have subspaces $\cF_{i-1}\in\cC_{i-1}=\bC_{i-1}$ and 
$\cF_{i+1}\in\cC_{i+1}=\bC_{i+1}$ satisfying $\cF_{i-1}\subsetneq \cF_i=\cU_i \subsetneq \cF_{i+1}.$  

$\boxed{(iii)\then (iv)}$ Assume that condition $(iii)$ holds and consider $1\leq i\leq r$ and a subspace $\cU_i\in\bC_i.$ By $(iii)$, we can find subspaces $\cU_{i-1}\in\bC_{i-1}$ and $\cU_{i+1}\in\bC_{i+1}$ such that $\cU_{i-1}\subsetneq\cU_i\subsetneq\cU_{i+1}.$ We apply the same argument to subspaces $\cU_{i-1}$ and $\cU_{i+1}$ in order to find subspaces $\cU_{i-2}\in\bC_{i-2}$ and $\cU_{i+2}\in\bC_{i+2}$ such that $\cU_{i-2}\subsetneq \cU_{i-1}$ and $\cU_{i+1}\subsetneq \cU_{i+2}$. This iterative process leads to a sequence of nested subspaces $\cU_1\subsetneq \dots\subsetneq \cU_i \subsetneq \dots \subsetneq \cU_r$ with $\cU_j\in\bC_j,$ for every $1\leq j\leq r.$ Then $(ii)$ follows by taking the flag $\cF=(\cU_1, \dots, \cU_i, \dots, \cU_r)$.

$\boxed{(iv)\then (i)}$ For each $i\in \{1,\dots , r\}$ and each subspace $\cU_i\in\bC_i$, there exist a flag $\cF^{(\cU_i)}\in\mathcal{F}_q((t_1,\dots, t_r),n)$ such that $\cF^{(\cU_i)}_i=\cU_i$ and $\cF^{(\cU_i)}_j\in\bC_j$, for all $1\leq j\leq r$. Therefore the flag code $\cC^{(i)}=\{\cF^{(\cU_i)}\ |\ \cU_i\in\bC_i\}$ satisfy that  $\cC^{(i)}_i=\bC_i$ and $\cC^{(i)}_j\subseteq \bC_j$, for all $1\leq j\leq r$. Now, consider the flag code $\cC=\bigcup_{i=1}^r \cC^{(i)}$. Clearly $\cC_i=\bC_i$ for all $1\leq i\leq r$ and then $\bC_1\times \dots\times \bC_r$ is a generating set of flag codes on $\bbF_q^n$ of type $(t_1,\dots, t_r)$.

%$(i)\then (ii)$ By hypotheses there exists a flag code $\cC\subseteq \mathcal{F}_q((t_1,\dots, t_r),n)$ such that $\cC_i=\bC_i$ for all $i=1,\dots, r$. Thus, $(ii)$ is true because of the definition of the projected codes (see (\ref{eq:def_proj})).

%$(ii)\then (iii)$ Consider any index $i\in\{1, \dots, r\}$ and take a subspace $\cU_i\in\bC_i$. By $(ii)$, there exists a flag $\cF\in\cF_q((t_1, \dots, t_r), n)$ such that $\cF_i=\cU_i$ and $\cF_{j}\in\bC_j$, for every $j=1, \dots, r$. If we put $\cF_0=\{0\}$ and $\cF_{r+1}=\bbF_q^n,$ it suffices to take subspaces $\cF_{i-1}\in\bC_{i-1}$ and $\cF_{i+1}\in\bC_{i+1}$ that satisfy condition $(iii)$.

%$(iv)\then (i)$ Assuming $(iv)$, it is clear that $\cC$ is a flag code generated by the sequence $\bC_1\times \dots\times \bC_r.$

\end{proof}

\begin{example}\label{ex: same generating set different flag codes}
Let $\{e_1, e_2, e_3\}$ be the standard $\bbF_q$-basis of $\bbF_q^3$ and consider the Cartesian product of constant dimension codes $\bC_1\times\bC_2$ with
$$
\bC_1 = \{ \langle e_1 \rangle, \langle e_2 \rangle\}\subset \cG_q(1,3) \quad \text{and} \quad
\bC_2 =\{ \langle  e_1, e_2 \rangle, \langle e_2, e_3 \rangle\}\subset \cG_q(2,3).
$$
Then $\cC=(\bC_1\times \bC_2)\cap \cF_q((1,2), 3)=\{\cF, \cF', \cF''\}$, where
$$
\cF= (\langle e_1 \rangle, \langle e_1, e_2 \rangle ), \quad 
\cF'= (\langle e_2 \rangle, \langle e_2, e_3 \rangle ), \quad \text{and} \quad
\cF'' = (\langle e_2 \rangle, \langle e_1, e_2 \rangle ).
$$
Since $\cC_i=\bC_i$, for $i=1,2$, it follows that $\bC_1\times\bC_2$ is a generating set of flag codes. 
In addition, we note that the flag code $\cC'=\{\cF, \cF'\}$ is also generated by $\bC_1 \times \bC_2$.
\end{example}

\begin{remark}
Clearly, every flag code $\cC\subseteq \mathcal{F}_q((t_1,\dots, t_r),n)$ is generated by the Cartesian product $\cC_1\times \dots\times \cC_r$ of its projected codes. However, this set of constant dimension codes can generate several different flag codes, as the previous example shows. Trying to avoid this situation, we focus on those flag codes such that the Cartesian product of their projected codes produces a unique flag code. 
\end{remark}

\begin{definition}
We say that a flag code $\cC\subseteq \mathcal{F}_q((t_1,\dots, t_r),n)$ is {\em determined by its projected codes} if it is the only flag code generated by the Cartesian product of its projected codes $\cC_1\times \dots\times \cC_r$. 
\end{definition}

In the following, we characterize flag codes determined by their projected codes. This problem is equivalent to characterize generating sets of constant dimension codes that just generate one flag code. 

As a first step, now we give the definition of a prominent flag code among all those generated by the same generating set of flag codes.

\begin{definition}\label{def: SIC}
A flag code $\cC$ of type $(t_1, \dots, t_r)$ on $\bbF_q^n$ is said to be \emph{subspace-inclusion-closed}  ({\em SIC}, for short) if, for every $1< i \leq r$ and every pair of flags $\cF, \cF'\in\cC$ such that $\cF_{i-1}\subset\cF'_{i}$, then the flag
$$
(\cF_1, \dots, \cF_{i-1}, \cF'_{i}, \dots, \cF'_r)
$$
lies on $\cC$.
\end{definition}

\begin{example}
Consider the flag codes $\cC$ and $\cC'$ in Example \ref{ex: same generating set different flag codes}. The flag code $\cC'$ is not SIC, since $\cF, \cF'\in\cC'$ satisfy $\cF'_1=\langle e_2\rangle \subset \cF_2=\langle e_1, e_2\rangle$ but the flag $\cF''=(\langle e_2\rangle,\langle e_1, e_2\rangle) \notin\cC'.$ On the other hand, $\cC$ is a SIC flag code.
\end{example}

Next we show that every generating set of flag codes produces a unique SIC flag code. We give its explicit expression and  prove that any other flag code generated by the same generating set is contained in this SIC flag code. 
\begin{theorem}\label{th:SIC}
Let $\bC_1\times \dots \times \bC_r\subseteq   \cG_q(t_1,n)\times \dots \times \cG_q(t_r,n)$ be a generating set of flag codes and consider the flag code
$$
\cC=(\bC_1\times\dots\times\bC_r)\cap \cF_q((t_1, \dots, t_r), n)\neq \emptyset .
$$
The following statements hold:
\begin{enumerate}[$(i)$]
\item $\cC$ is a SIC flag code generated by $\bC_1\times \dots \times \bC_r$.
\item If $\cC'$ is another flag code generated by $\bC_1\times \dots \times \bC_r$, then $\cC'\subseteq \cC$. Moreover, the equality holds if, and only if, $\cC'$ is SIC.
\end{enumerate}
 In particular, the flag code $\cC$ is the unique SIC flag code generated by $\bC_1\times \dots \times \bC_r$.

\end{theorem}
\begin{proof}
$(i)$ By means of Theorem \ref{theo:carac_seqgen}, we have $\cC\neq \emptyset$ and $\cC_i=\bC_i,$ for every $1\leq i\leq r,$ i.e., $\cC$ is a flag code generated by $\bC_1\times \dots\times\bC_r$. Now consider an index $1<i\leq r$ and suppose that there exist different flags $\cF, \cF'\in\cC$ such that $\cF_{i-1}\subset \cF'_i.$ Then the sequence $(\cF_1, \dots, \cF_{i-1}, \cF'_i, \dots, \cF'_r)$ forms a flag (is an element in $\cF_q((t_1, \dots, t_r), n)$) and, of course, is an element in the Cartesian product $\bC_1\times \dots\times\bC_r$. Consequently, it is a flag in $\cC$ and then $\cC$ is SIC.

$(ii)$ If $\cC'$ is another flag code generated by $\bC_1\times \dots \times \bC_r$, then we trivially obtain that $\cC'\subseteq \cC$ because of the definition of $\cC$. Moreover, if $\cC'=\cC$, then they both are SIC flag codes. For the converse, assume that $\cC'$ is a SIC flag code generated by $\bC_1\times \dots \times \bC_r$. Let us see that the other inclusion $\cC\subseteq \cC'$ holds too.  Given a flag $\cF=(\cF_1, \dots, \cF_r)\in\cC$, we show that, for every $1\leq i\leq r$, we can find a flag in $\cC'$ having $\cF_1, \dots, \cF_i$ as its subspaces. We argue by induction on $i$. For $i=1$, clearly, $\cC'$ contains flags with $\cF_1$ as their first subspace since $\cF_1\in\cC_1=\bC_1=\cC'_1$. Now, take $1< i \leq r$ and assume that there exists a flag $\cF'\in\cC'$ with $\cF'_j=\cF_j$ for $1\leq j\leq i-1$. We will prove that the result also holds for $i$. Since $\cF_{i}\in\cC_{i}=\bC_{i}=\cC'_{i}$, we can find a flag $\cF''\in \cC'$ with $\cF''_{i}=\cF_{i}$. Thus, the flags $\cF',\cF''\in \cC'$ satisfy $\cF'_{i-1}=\cF_{i-1}\subset \cF_{i} = \cF''_{i}.$
Then, since $\cC'$ is SIC, we conclude that the flag 
$$
(\cF_1, \dots, \cF_{i-1}, \cF_{i},  \cF''_{i+1},\dots, \cF''_r) \in\cC'.
$$
In particular, the case $i=r$ proves that $\cF\in\cC'$. As a consequence, we have $\cC=\cC'$ and then the uniqueness of the SIC flag code $\cC.$ 
\end{proof}

This result, in particular, leads to the next characterization for SIC flag codes.

\begin{corollary}\label{cor: char SIC}
    Let $\cC$ be a flag code of type $(t_1, \dots, t_r)$ on $\bbF_q^n$ and consider its projected codes $\cC_1, \dots, \cC_r.$ The following statements are equivalent:
    \begin{enumerate}[$(i)$]
    \item $\cC$ is a SIC flag code,
    \item$(\cC_1\times\dots\times\cC_r)\cap\cF_q((t_1, \dots, t_r), n)=\cC$, 
    \item $(\cC_1\times\dots\times\cC_r)\cap\cF_q((t_1, \dots, t_r), n)\subseteq \cC.$
    \end{enumerate}
\end{corollary}

\medskip

Now, we introduce the second tool we need in order to characterize the flag codes that are determined by their projected codes. 

\begin{definition}
Given a flag code $\cC\subseteq  \mathcal{F}_q((t_1,\dots, t_r),n)$ and a flag $\cF\in \cC$, for every $1\leq i\leq r$, we define the {\em multiplicity  of the subspace $\cF_i$ with respect to $\cC$}, denoted by $m_{\cC}(\cF_i)$, as the number of times that this subspace appears as the $i$-th subspace of different flags of $\cC$.
\end{definition}

\begin{example}\label{ex:mult}
Consider the flag code $\cC$ of type $(1, 2)$ on $\bbF_q^3$ given in Example \ref{ex: same generating set different flag codes}.  We have that $m_{\cC}(\cF'_1)=m_{\cC}(\cF''_1)=m_{\cC}(\cF_2)=m_{\cC}(\cF''_2)=2$, whereas $m_{\cC}(\cF_1)=m_{\cC}(\cF'_2)=1$. 
\end{example}

We are now in conditions to characterize sets of constant dimension codes $\bC_1\times \dots\times \bC_r$ producing a single flag code.
Recall that, by means of Theorem \ref{th:SIC}, a generating set of flag codes always provides a SIC flag code $\cC$, which contains all the flag codes generated by the same family of constant dimension codes. Hence, we  just need to ensure that the unique SIC flag code generated by  $\bC_1\times \dots\times \bC_r$ does not admit any subcode generated by the same set of constant dimension codes.

\begin{theorem}\label{th:SICunic}
Let $\cC$ be the SIC flag code generated by the sequence of constant dimension codes $\bC_1\times \dots \times \bC_r\subseteq   \cG_q(t_1,n)\times \dots \times \cG_q(t_r,n)$. The following statements are equivalent:
\begin{enumerate}[$(i)$]
\item $\cC$ is the only flag code generated by $\bC_1\times \dots \times \bC_r$.
\item For each $\cF\in\cC$, there exists $i\in\{1,\dots,r\}$ such that $m_{\cC}(\cF_i)=1$.
\end{enumerate}
\end{theorem}
\begin{proof}
 $ $\newline
 
$\boxed{(i)\then (ii)}$ Assume that there is some flag $\cF\in\cC$ such that $m_{\cC}(\cF_i)>1$ for every $1\leq i\leq r$. Then the code $\cC\setminus \{\cF\}$ is different from $\cC$ and it is also generated by $\bC_1\times \dots \times \bC_r$.

$\boxed{(ii)\then (i)}$  Suppose now that there is another flag code $\cC'\neq \cC$ generated by the set of constant dimension codes $\bC_1\times \dots \times \bC_r$. By means of Theorem \ref{th:SIC}, it follows that $\cC'\subseteq\cC$. Since the codes are different, we can find
 $\cF\in\cC$ such that $\cC'\subseteq \cC\setminus\{\cF\}\subseteq \cC$. In this case, we have
$$
\cC_i=\bC_i=\cC'_i\subseteq \left(\cC\setminus\{\cF\}\right)_i \subseteq\cC_i,
$$
that is, $\cC_i=\bC_i=\left(\cC\setminus\{\cF\}\right)_i$, for every $1\leq i\leq r$. This means that every subspace $\cF_i$ in the flag $\cF$ still appears in others flags in $\cC\setminus\{\cF\}$. In terms of multiplicities, this means that $m_{\cC}(\cF_i)\geq 2$, for every $1\leq i\leq r$, which is a contradiction.

\end{proof}

\begin{corollary}\label{cor:Cdeter_by_proj}
A flag code $\cC$ is determined by its projected codes if, and only if, it is a SIC flag code and each flag of $\cC$ has a subspace with multiplicity $1$ with respect to $\cC$.
\end{corollary}
\begin{proof}
If $\cC$ is determined by its projected codes, then it is the only flag code generated by the Cartesian product $\cC_1\times\dots\times\cC_r$. Thus, by Theorem \ref{th:SIC}, $\cC$ is the SIC flag code $\cC=(\cC_1\times\dots\times\cC_r)\cap\cF_q((t_1, \dots, t_r), n)$. Now, Theorem \ref{th:SICunic} ensures that each flag of $\cC$ has a subspace with multiplicity $1$ with respect to $\cC$. The converse follows directly from Theorem \ref{th:SICunic}.
\end{proof}

\begin{example}\label{ex:Cnotdeterm}
The flag code $\cC$ in Example \ref{ex: same generating set different flag codes} is SIC but $m_{\cC}(\cF''_1)=m_{\cC}(\cF''_2)=2$. Consequently, $\cC$  is not  determined by their projected codes. In fact, this sequence of constant dimension codes also generates the flag code $\cC'=\{\cF,\cF'\}$.  In contrast, the flag code $\cC''\subseteq\cF_q((1,2),5)$ consisting of flags 
$$
\begin{array}{rcl}
\cF^1 & = & (\left\langle  e_1 \right\rangle, \left\langle e_1, e_2 \right\rangle), \\
\cF^2 & = & (\left\langle  e_2 \right\rangle, \left\langle e_1, e_2 \right\rangle), \\
\cF^3 & = & (\left\langle  e_3 \right\rangle, \left\langle e_3, e_4 \right\rangle), \\
\cF^4 & = & (\left\langle  e_3 \right\rangle, \left\langle e_3, e_5 \right\rangle), \\
\end{array}
$$
is determined by its projected codes. In other words, it is the only flag code that can be generated by the Cartesian product $\cC''_1\times\cC''_2$. 
\end{example}

%%%%%%%%%%%%%%%%%%%%%%%%%%%%%%%%%%%%%%%%%%%%

\section{Equivalence for flag codes in terms of their projected codes}\label{sec: l and sl eq}

In this section, we generalize the notion of (semi)linear equivalence given in Section \ref{subsec:constdimcod} for constant dimension codes (see Definition \ref{def:const_equiv}) to the flag codes scenario. We focus on the relationship between the (semi)linear equivalence of flag codes and the one of their corresponding projected codes. 
In particular, we give sufficient conditions for flag codes to be (semi)linearly equivalent if, and only if,  their projected codes are (semi)linearly equivalent. As a consequence of these questions, we finish the section with the computation of the linear and semilinear automorphism groups of SIC flag codes in terms of the ones of their projected codes.

We start considering the actions of $\GL_n(\bbF_q)$ and $\GaL_n(\bbF_q)=\GL_n(\bbF_q) \rtimes \mathrm{Aut}(\bbF_q)$ on $\cF_q((t_1, \dots, t_r), n)$, induced by their actions on the corresponding  Grassmann varieties (see (\ref{eq:acGL}) and (\ref{eq:acGaL})). More precisely, given a flag $\cF=(\cF_1, \dots, \cF_r)$ of type $(t_1, \dots, t_r)$ on $\bbF_q^n$ and elements $A\in\GL_n(\bbF_q)$ and $\varphi \in\mathrm{Aut}(\bbF_q)$, the operations
$$
\cF\cdot A = (\cF_1\cdot A, \dots, \cF_r \cdot A) \ \text{and} \ \cF\cdot (A, \varphi) = (\cF_1\cdot (A, \varphi), \dots, \cF_r \cdot (A, \varphi))
$$
define respective actions of $\GL_n(\bbF_q)$ and $\GaL_n(\bbF_q)$ on flags on $\bbF_q^n$ of type $(t_1, \dots, t_r)$.

As a result, given a flag code $\cC\subseteq \cF_q((t_1, \dots, t_r), n)$, for each $A\in\GL_n(\bbF_q)$ and $\varphi \in\mathrm{Aut}(\bbF_q)$, the sets
\[
\cC\cdot A=\{\cF\cdot A\ |\  \cF\in \cC\} \quad \mbox{ and } \quad \cC\cdot (A,\varphi)=\{\cF\cdot (A,\varphi)\ |\  \cF\in \cC\}.
\]
are flag codes in $\cF_q((t_1, \dots, t_r), n).$

As for constant dimension codes, these actions preserve cardinality. Moreover, the flag distance is also preserved, since it is defined as a sum of subspace distances. For this reason, they lead us to consider the following definitions of equivalence for flag codes.

\begin{definition}
Two flag codes $\cC$ and $\cC'$ of type $(t_1, \dots, t_r)$ on $\bbF_q^n$ are said to be \emph{linearly equivalent}  (resp. {\em semilinearly equivalent}) if there exists an element $A\in\GL_n(\bbF_q)$ (resp. $(A,\varphi)\in \GaL_n(\bbF_q)$) such that $\cC'=\cC\cdot A$ (resp. $\cC'=\cC\cdot (A,\varphi)$).

The orbit of $\cC$ under the action of $\GL_n(\bbF_q)$ on the flag variety is the set
$$
\cC\cdot\GL_n(\bbF_q) = \{ \cC\cdot A \ | \ A\in\GL_n(\bbF_q)\}
$$
containing all the flag codes that are linearly equivalent to $\cC$. Similarly, the set
$$
\cC\cdot\GaL_n(\bbF_q) = \{ \cC\cdot (A, \varphi) \ | \ (A, \varphi)\in\GaL_n(\bbF_q) \}
$$
 is the equivalence class of $\cC$ under the semilinearly equivalence relation. Finally, just as for constant dimension codes (see Definition \ref{def:aut_constdim}), we consider the corresponding stabilizers of these actions. Given a flag code $\cC$, the {\em linear automorphism group} of $\cC$, is the set  
\[
\Aut(\cC)=\{A\in  \GL_n(\bbF_q)\ |\ \cC\cdot A=\cC\}
\]
and its {\em semilinear automorphism group} is defined as 
\[
\SAut(\cC)=\{(A,\varphi)\in \GaL_n(\bbF_q)\ |\ \cC\cdot (A,\varphi)=\cC\}.
\]
\end{definition}

\begin{remark}\label{remark:equiv_impl_proj_equiv}
Notice that if $\cC, \cC'$ are linearly (resp. semilinearly) equivalent flag codes, then $\cC'=\cC\cdot A$, for some $A\in \GL_n(\bbF_q)$ (resp. $\cC'=\cC \cdot (A, \varphi)$, for some $(A, \varphi)\in\GaL_n(\bbF_q)$). Consequently, the corresponding projected codes satisfy that 
\begin{equation}\label{eq:proj_equiv}
\cC'_i=(\cC\cdot A)_i=\cC_i\cdot A
\end{equation}
(resp. $\cC'_i=(\cC \cdot (A, \varphi))_i=\cC_i \cdot (A, \varphi)$),
 for all $1\leq i\leq r$. In other words, the projected codes of (semi)linearly equivalent flag codes are (semi)linearly equivalent constant dimension codes. Moreover, all of them are related by the same group element in $\GL_n(\bbF_q)$ (resp. $\GaL_n(\bbF_q)$). The converse of this fact is not true in general, as we can see in the following example.
\end{remark}

\begin{example}\label{ex:reciprocno}
Consider the standard $\bbF_q$-basis $\{e_1, e_2, e_3\}$ of $\bbF_q^3$ and the following flags 
$$
\begin{array}{ll}
 \cF^1 = (\langle e_1 \rangle, \langle e_1, e_2 \rangle ), \quad & \cF^2= (\langle e_2 \rangle, \langle e_2, e_3 \rangle ), \\
\cF^3 = (\langle e_2 \rangle, \langle e_1, e_2 \rangle ), \quad & \cF^4= (\langle e_1 \rangle, \langle e_1, e_3 \rangle )
\end{array}
$$
of type $(1,2)$ on $\bbF_q^3$. Now we form two different flag codes $\cC=\{\cF^1, \cF^2, \cF^3\}$ and
$\cC'=\{ \cF^3, \cF^4\}$. With this notation, we have projected codes
$$
\cC_1= \cC'_1= \{\langle e_1 \rangle, \langle e_2 \rangle\},  \quad \cC_2=\{ \langle e_1, e_2 \rangle, \langle e_2, e_3 \rangle )\}, \quad   \cC'_2=\{ \langle e_1, e_2 \rangle, \langle e_1, e_3 \rangle )\}.
$$
If we put
$$
A
=
\begin{pmatrix}
0 & 1 & 0\\
1 & 0 & 0\\
0 & 0 & 1
\end{pmatrix}
\in\GL_3(\bbF_q),
$$
then it is clear that $\cC_1=\cC'_1\cdot A$ and $\cC_2=\cC'_2\cdot A.$ However, $\cC$ and $\cC'$ cannot be (semi)linearly equivalent since $|\cC|\neq|\cC'|.$ 

\end{example}

The following question  naturally arises: for what family of flag codes does the converse of the property explained in Remark \ref{remark:equiv_impl_proj_equiv} hold true? That is, we are interested in those flag codes such that the equivalence (both linear or semilinear) of the corresponding projected codes (with the same element of $\GL_n(\bbF_q)$ or  $\GaL_n(\bbF_q)$) ensures the equivalence (linear or semilinear) between the initial flag codes. In light of the previous example, it is not enough to think of SIC flag codes: the code $\cC$ in Example \ref{ex:reciprocno} is SIC but we can find non-equivalent flag codes to $\cC$ (as $\cC'$ in the same example) having equivalent projected codes.  Fortunately, we will see that the flag codes determined by their projected codes fulfill the desired property. To this end, we now show that both properties of being SIC and being determined by its projected codes are invariant under (semi)linear equivalence.

\begin{theorem}\label{theo:deter}
Let $\cC$ be a flag code of $\cF_q((t_1, \dots, t_r), n)$. 
\begin{enumerate}[$(i)$]
\item  The flag code $\cC$ is SIC if, and only if, every (semi)linearly equivalent flag code to $\cC$ is also SIC.
\item $\cC$ is determined by its projected codes if, and only if, every (semi)linearly equivalent flag code to $\cC$ is also determined by its projected codes. 
\end{enumerate}
\end{theorem}
\begin{proof}
$(i)$ According to Corollary \ref{cor: char SIC}, the flag code $\cC$ is SIC if, and only if, 
\begin{equation}\label{eq: sic condition}
(\cC_1\times\dots\times\cC_r)\cap\cF_q((t_1, \dots, t_r), n)=\cC 
\end{equation}
Now, for every element $A\in\GL_n(\bbF_q)$, the equality $(\ref{eq: sic condition})$ holds if, and only if, 
$$
(\cC_1\cdot A\times\dots\times\cC_r\cdot A)\cap\cF_q((t_1, \dots, t_r), n)=\cC\cdot A.
$$
Thus, $\cC$ is SIC if, and only if, the flag code $\cC\cdot A$ is SIC too. We can argue analogously for semilinear equivalence.

$(ii)$ Notice that, for every $A\in \GL_n(\bbF_q)$, the flag code $\cC\cdot A$ is generated by the Cartesian product of constant dimension codes $\cC_1\cdot A \times \dots \times \cC_r\cdot A$ by (\ref{eq:proj_equiv}). Moreover,  for each $\cF\in\cC$, one has that 
\[
m_{\cC\cdot A}((\cF\cdot A)_i)=m_{\cC\cdot A}(\cF_i\cdot A)=m_{\cC}(\cF_i),
\] for all $1\leq i\leq r$. Consequently, by statement $(i)$ and Corollary  \ref{cor:Cdeter_by_proj}, we obtain that $\cC\cdot A$ is determined by its projected codes  if, and only if $\cC$ satisfies the same condition.
The proof for semilinearly equivalence runs analogously. 
\end{proof}

Now, as a direct consequence of this theorem, we obtain the converse part of Remark  \ref{remark:equiv_impl_proj_equiv}.

\begin{corollary}\label{cor:deter}
Let $\cC$ be a flag code in $\cF_q((t_1, \dots, t_r), n)$ determined by its projected codes and assume that there exist $\cC'\subseteq \cF_q((t_1, \dots, t_r), n)$ and $A\in \GL_n(\bbF_q)$ (resp. $(A,\varphi)\in \GaL_n(\bbF_q)$) such that $\cC'_i=\cC_i\cdot A$ (resp. $\cC_i'=\cC_i\cdot (A,\varphi)$), for all $1\leq i\leq r$. Then $\cC'=\cC\cdot A$ (resp. $\cC'=\cC\cdot (A,\varphi)$).
\end{corollary}
\begin{proof}
Given $A\in \GL_n(\bbF_q)$, by Theorem \ref{theo:deter} (part $(ii)$), the flag code $\cC\cdot A$ is determined by its projected codes. Since $(\cC\cdot A)_i=\cC_i\cdot A=\cC_i'$, both flag codes $\cC\cdot A$ and $\cC'$ are generated by the Cartesian product of constant dimension codes $\cC_1\cdot A\times \dots \times \cC_r\cdot A$ and then necessarily $\cC\cdot A=\cC'$. The proof for a $(A,\varphi)\in \GaL_n(\bbF_q)$ is analogous. 
\end{proof}

\begin{remark}
Notice that the condition of being determined by the projected codes in the previous corollary is needed in the following sense. If a flag code $\cC\subseteq \cF_q((t_1, \dots, t_r), n)$ is not determined by its projected codes, then there exists, at least, another flag code $\bar{\cC}$ also generated by the set $\cC_1\times\dots\times\cC_r$. Moreover, we can always find $\bar{\cC}$ having size $|\bar{\cC}|\neq|\cC|,$ it suffices to take into account the following possibilities:
\begin{itemize}
    \item If $\cC$ is the SIC flag code generated by $\cC_1\times\dots\times\cC_r$, just take another flag code $\bar{\cC}$ generated by $\cC_1\times\dots\times\cC_r$, which is a proper subset of $\cC.$
    \item Otherwise, just consider $\bar{\cC}$ as the SIC flag code generated by $\cC_1\times\dots\times\cC_r$, which satisfies $\cC\subsetneq \bar{\cC}.$
\end{itemize}
In any case, there exists a flag code $\bar{\cC}$ with projected codes $\bar{\cC}_i=\cC_i$ for all $1\leq i\leq r$ and satisfying $|\bar{\cC}|\neq|\cC|.$ After this remark, it is clear that, for every $A\in \GL_n(\bbF_q)$, the code $\cC'=\bar{\cC}\cdot A$ satisfies $\cC'_i= \bar{\cC}_i\cdot A= \cC_i\cdot A$ for every $1\leq i\leq r$. However, $\cC$ and $\cC'$ cannot be linearly equivalent, since $|\cC'|=|\bar{\cC}|\neq|\cC|.$ The same argument holds for semilinearly equivalence.
\end{remark}

To conclude our study on the relationship between (semi)linearly  equivalent flag codes and their corresponding projected codes, we focus now on the (semi)linear automorphism group of a flag code in terms of those of its projected codes. Clearly, given a flag code $\cC\subseteq \cF_q((t_1, \dots, t_r), n)$, one always has the inclusions

\begin{equation}\label{eq: aut inclusion}
    \Aut(\cC)\subseteq \bigcap_{i=1}^r \Aut(\cC_i) \quad \text{and} \quad \SAut(\cC)\subseteq \bigcap_{i=1}^r \SAut(\cC_i).
\end{equation}
However, these inclusions can be strict, as we can see in the next example.

\begin{example}\label{ex:inclusioAut}
Let  $\{e_1, e_2, e_3\}$ be the standard $\bbF_q$-basis of $\bbF_q^3$ and  consider the following flags 
$$
 \cF^1 = (\langle e_1 \rangle, \langle e_1, e_2 \rangle ), \quad \cF^2= (\langle e_2 \rangle, \langle e_2, e_3 \rangle ) \quad \text{and} \quad \cF^3 = (\langle e_3 \rangle, \langle e_1, e_3 \rangle ).
$$
of type $(1,2)$ on $\bbF_q^3$. The flag code $\cC=\{\cF^1, \cF^2, \cF^3\}$ has projected codes
$$
\cC_1= \{\langle e_1 \rangle,  \langle e_2 \rangle, \langle e_3\rangle\} \quad \text{and} \ \cC_2=\{ \langle e_1, e_2 \rangle, \langle e_2, e_3 \rangle ), \langle e_1, e_3 \rangle\}.
$$
Moreover, the matrix
$$
A
=
\begin{pmatrix}
0 & 1 & 0\\
1 & 0 & 0\\
0 & 0 & 1
\end{pmatrix}
\in\GL_3(\bbF_q),
$$
satisfies  $\cC_i=\cC_i\cdot A$ for $i=1,2,$ that is, $A\in\Aut(\cC_1)\cap \Aut(\cC_2)$. However, $A\notin \Aut(\cC),$ since $\cF^2\cdot A= (\langle e_1 \rangle, \langle e_1, e_3 \rangle )\notin \cC$ and $\cF^3\cdot A= (\langle e_3 \rangle, \langle e_2, e_3 \rangle )\notin \cC$.
\end{example}

The code $\cC$ in the previous example is not SIC as it does not contain the flags
$$
(\langle e_1 \rangle, \langle e_1, e_3 \rangle) \quad \text{and} \quad  (\langle e_3 \rangle, \langle e_2, e_3 \rangle).
$$
Notice that these are precisely the flags making $\cC\neq \cC\cdot A$. This is no coincidence: the next result proves that the inclusions in (\ref{eq: aut inclusion}) become equalities for SIC flag codes.

\begin{theorem}\label{theo:igualaut}
Let $\cC$ be a SIC flag code in $\cF_q((t_1, \dots, t_r), n)$. Then
\[
\Aut(\cC)=\bigcap_{i=1}^r \Aut(\cC_i) \quad \text{and} \quad \SAut(\cC)=\bigcap_{i=1}^r \SAut(\cC_i).
\]
\end{theorem}
\begin{proof}
 By (\ref{eq:proj_equiv}), given $A\in \bigcap_{i=1}^r \Aut(\cC_i)$,  the flag code $\cC\cdot A$  is generated by the set of constant dimension codes $\cC_1\cdot A \times \dots \times \cC_r\cdot A=\cC_1\times \dots \times \cC_r$. 
 Therefore, by means of Theorem \ref{th:SIC} (part $(ii)$), we have $\cC\cdot A\subseteq \cC$.   
 Moreover, since $|\cC\cdot A|=|\cC|$,  we conclude $\cC=\cC\cdot A$ and then $A\in \Aut(\cC)$. The proof for the semilinear automorphism group runs analogously.
\end{proof}

\begin{remark}
As we have seen in Example \ref{ex:reciprocno}, the SIC property is not sufficient to deduce the (semi)linear equivalence of flag codes from the (semi)linear equivalence of the corresponding projected codes. To obtain this, we need the flag code to satisfy the stronger property of being determined by its projected codes  (Corollary \ref{cor:deter}). In contrast, Theorem \ref{theo:igualaut} shows that for the  (semi)linear automorphism group  is is enough for the flag code to verify the weakest property of being SIC.
\end{remark}

We end this section with an example showing that the previous result does not characterizes those flag codes for which (\ref{eq: aut inclusion}) becomes an equality:  there are flag codes satisfying this equality but not being SIC, as the next example illustrates.
 
\begin{example}
Let  $\{e_1, e_2, e_3\}$ be the standard $\bbF_q$-basis of $\bbF_q^3$ and  consider the flag code $\cC=\{\cF^1, \cF^2\}$ with 
    $$
    \cF^1 = (\langle e_1 \rangle, \langle e_1, e_2 \rangle ), \quad \cF^2= (\langle e_2 \rangle, \langle e_2, e_3 \rangle).
    $$
   Since the flag $(\langle e_2 \rangle, \langle e_1, e_2 \rangle )\notin \cC$, it follows that $\cC$ is not SIC. Concerning its linear automorphism group, we have 
    $$
    \Aut(\cC)=\left\lbrace
\begin{pmatrix}
a & 0 & 0\\
0 & b & 0\\
0 & d & c
\end{pmatrix}, 
\ \   a,b,c\in\bbF_q^\ast,\  d\in\bbF_q \ 
 \right\rbrace
 .
$$
Moreover, we it is easy to check that
$$
\begin{array}{ccl}
\Aut(\cC_1) & = & 
\left\lbrace 
\begin{pmatrix}
a & 0 & 0\\
0 & b & 0\\
e & d & c
\end{pmatrix},
\begin{pmatrix}
0 & a & 0\\
b & 0 & 0\\
e & d & c
\end{pmatrix},
\ \ a,b,c\in\bbF_q^\ast, \ d, e\in\bbF_q
\right\rbrace, \\
\Aut(\cC_2) & = & \left\lbrace \begin{pmatrix}
a & f & 0\\
0 & b & 0\\
0 & d & c
\end{pmatrix},
\begin{pmatrix}
0 & f & a\\
0 & b & 0\\
c & d & 0
\end{pmatrix},
\ \ a,b,c\in\bbF_q^\ast, \ d,f\in\bbF_q
\right\rbrace
\end{array}
$$
and conclude that $\Aut(\cC)= \Aut(\cC_1)\cap\Aut(\cC_2)$.
\end{example}

\section{Examples of flag codes determined by their projected codes}\label{sec: examples}

This section is devoted to the study of specific families of flag codes determined by their projected codes. We start by defining two new classes of flag codes: the ones of \emph{increasing} and \emph{decreasing} flag codes and we show that they are examples of flag codes determined by their projected codes. Moreover, we prove that, for many choices of the type vector $(t_1, \dots, t_r)$, increasing/decreasing flag codes arise when considering flag codes with projected codes of maximum distance and, in particular, optimum distance flag codes.

Given a flag code $\cC$, for every $1<i\leq r$, its projected codes $\cC_{i-1}$ and $\cC_i$ are always related by the subspace inclusion: any subspace in $\cC_{i-1}$ is contained in, at least, one subspace in $\cC_i$. Conversely, any subspace in $\cC_i$ contains, at least, one subspace in $\cC_{i-1}$. In the following definition we name those flag codes for which these inclusion relations between the subspaces of $\cC_{i-1}$ and $\cC_i$ define maps (more precisely, surjective maps). 

\begin{definition}\label{def:incdec}
A flag code $\cC$ of type $(t_1, \dots, t_r)$ on $\bbF_q^n$ is said to be \emph{increasing} if, for every $1< i \leq r$ and every subspace $\cV\in\cC_i$, there is just a single subspace $\cU\in\cC_{i-1}$ such that $\cU\subset \cV$. In other words, $\cC$ is increasing if it is possible to define the map   $\alpha_i: \cC_i \en \cC_{i-1} $, mapping every subspace $\cV\in\cC_i$ into $\alpha_i(\cV)=\cU\in\cC_{i-1}$ such that $\cU\subset \cV$:

\begin{equation}\label{eq: map increasing}
\begin{array}{ccccc}
\alpha_i & : & \cC_i & \longrightarrow & \cC_{i-1}\\
           &   &  \cV  & \longmapsto     & \alpha_i(\cV)=\cU,
\end{array}
\end{equation}
for every $1< i \leq r$.

Similarly, $\cC$ is called \emph{decreasing} if, for every  $1 < i \leq r$ and every subspace $\cU\in\cC_{i-1}$  there is a single subspace $\cV\in\cC_{i}$ containing $\cU$. In other words, $\cC$ is decreasing if it is possible to define the map $\beta_i:\cC_{i-1} \en \cC_{i}$, mapping every subspace $\cU\in\cC_{i-1}$ into the only subspace $\cV\in\cC_{i}$ containing $\cU$:

\begin{equation}\label{eq: map decreasing}
\begin{array}{ccccc}
\beta_i & : & \cC_{i-1} & \longrightarrow & \cC_{i}\\
           &   &  \cU  & \longmapsto     & \beta_i(\cU)=\cV,
\end{array}
\end{equation}
for every $1< i \leq r$.

\end{definition}

\begin{remark}\label{rem:alpha}
Following the notation of Definition \ref{def:incdec}, it is useful to note that if the map  $\alpha_i$  defined in (\ref{eq: map increasing}) exists for some index $1 < i \leq r$, then it is always a surjective map and then $| \cC_{i-1}|\leq | \cC_{i}|$. Moreover, if one has that $| \cC_{i-1}|= | \cC_{i}|$ and the map $\alpha_i$ exists, then it is a bijection and its inverse map is precisely the map $\beta_i$ defined in (\ref{eq: map decreasing}).
\end{remark}

\begin{example}
Consider the standard $\bbF_q$-basis $\{e_1, e_2, e_3, e_4\}$ of $\bbF_q^4$ and form the flags
$$
\begin{array}{ccc}
\cF^1 & = & (\langle e_1 \rangle, \langle e_1, e_2 \rangle,  \langle e_1, e_2, e_3 \rangle ),\\
\cF^2 & = & (\langle e_1 \rangle, \langle e_1, e_2 \rangle, \langle e_1, e_2, e_4 \rangle ),\\
\cF^3 & = & (\langle e_3 \rangle, \langle e_3, e_4 \rangle, \langle e_1, e_3, e_4 \rangle ).
\end{array}
$$
The code $\cC=\{\cF^1, \cF^2, \cF^3\}$ is increasing but it is not decreasing since the subspace $\langle e_1, e_2 \rangle\in\cC_2$ is contained in two different subspaces $\langle e_1, e_2, e_3 \rangle \neq \langle e_1, e_2, e_4 \rangle $ of $\cC_3$.
\end{example}

The name given to the families defined in Definition \ref{def:incdec} comes from the following result. 

\begin{proposition}\label{prop: cardinality increasing}
Let $\cC$ be  an increasing (resp. decreasing) flag code of type $(t_1, \dots, t_r)$ on $\bbF_q^n$. Then 
$$
|\cC_1| \leq \dots \leq |\cC_r|=|\cC| \ \ \left( \text{resp.} \ \ |\cC|=|\cC_1| \geq \dots \geq |\cC_r|\right).
$$
In other words, the cardinalities of the projected codes of an increasing (resp. decreasing) flag code code form an increasing (resp. decreasing) sequence of positive integers.
\end{proposition}
\begin{proof}
Since $\cC$ is increasing, we can define the maps $\alpha_i$ given in  (\ref{eq: map increasing}), for all $1<i\leq r$. Notice that these are surjective maps. Therefore $|\cC_1| \leq \dots \leq |\cC_r|$.

Let us see that $|\cC_r|=|\cC|.$ By the definition of projected code $|\cC_r|\leq |\cC|$, so let us assume that $|\cC_r|<|\cC|$. In this case, we can find different flags $\cF, \cF'\in\cC$ such that $\cF_r=\cF'_r$. Since $\cF\neq\cF'$, they must differ in, at least, a subspace. Take $1\leq j < r$ the maximum index such that $\cF_j\neq\cF'_j$. In this case, we find a subspace $\cF_{j+1}=\cF'_{j+1}\in\cC_{j+1}$ containing different subspaces $\cF_j\neq\cF'_j\in\cC_j$, which is a contradiction with the increasing condition. Thus,  $|\cC_r|=|\cC|$.

The proof is analogous for decreasing flag codes.
\end{proof}

Notice that, by means of Proposition \ref{prop: cardinality increasing} a flag code being at the same time increasing and decreasing is, in particular, disjoint (see Section  \ref{sec:backgroundFlags}). However, the converse is not true as we can see in the next example.

\begin{example}\label{ex: disjoint but not inc dec}
Let $\{e_1, \dots, e_6\}$ denote the standard $\bbF_q$-basis of $\bbF_q^6$ and consider the flag code $\cC=\{\cF, \cF'\}$ of type $(2,4)$ on $\bbF_q^6$, where
$$
\begin{array}{ccccc}
\cF  &=& (\cF_1, \cF_2)   &=& (\langle e_1, e_2 \rangle, \langle e_1, e_2, e_3, e_4 \rangle)\\
\cF' &=& (\cF'_1, \cF'_2) &=& (\langle e_3, e_4 \rangle, \langle e_3, e_4, e_5, e_6 \rangle).
\end{array}
$$
The flag code $\cC$ is disjoint but it is neither increasing ($\cF_2$ contains both $\cF_1\neq\cF'_1$) nor decreasing ($\cF'_1$ is contained in both $\cF_2\neq\cF'_2$).
\end{example}

On the other hand, the disjointness condition for flag codes relates the ones of being increasing and decreasing. 

\begin{theorem}\label{th: disjoint inc iff dec}
Let $\cC$ be a disjoint flag code of type $(t_1, \dots, t_r)$ on $\bbF_q^n$. Then $\cC$ is increasing if, and only if, it is decreasing. 
\end{theorem}
\begin{proof}
Let $\cC$ be a disjoint flag code, i.e., $|\cC_1|=\dots=|\cC_r|=|\cC|$ and assume that $\cC$ is increasing. Then, the maps $\alpha_i$ given in  (\ref{eq: map increasing}) exist for all $1<i\leq r$. Now, Remark \ref{rem:alpha} states the existence of  the maps $\beta_i$ defined in (\ref{eq: map decreasing}), for all $1<i\leq r$. Thus, $\cC$ is decreasing. 

For the converse, it suffices to apply analogous arguments with the maps $\beta_i$ defined in (\ref{eq: map decreasing}).
\end{proof}

Here below we see that being increasing (or decreasing) has implications on the multiplicities of the subspaces of the corresponding flags of a given flag code.

\begin{proposition}\label{prop: multiplicity increasing}
Let $\cC$ be  an increasing (resp. decreasing) flag code of type $(t_1, \dots, t_r)$ on $\bbF_q^n$. For every flag $\cF\in\cC$, it holds
$$
m_\cC(\cF_1)\geq \dots \geq m_\cC(\cF_r)=1 \quad \ (\text{resp.} \ 1=m_\cC(\cF_1)\leq \dots \leq m_\cC(\cF_r)).
$$
\end{proposition}
\begin{proof}
Assume that $\cC$ is an increasing flag code. Then, for every flag $\cF\in\cC$ and $1< i\leq r$, the subspace $\cF_i$ only contains $\cF_{i-1}\in\cC_{i-1}$. This means that $\cF_i$ can only be part of those flags in $\cC$ in which $\cF_{i-1}$ also appears. Thus $m_\cC(\cF_{i-1})\geq m_\cC(\cF_i)$. Moreover, since $|\cC|=|\cC_r|$, we conclude that $m_\cC(\cF_r)=1.$ The proof for decreasing flag codes is analogous.
\end{proof}

Next, we prove that increasing (resp. decreasing) flag codes are closed for the subspace inclusion (SIC).
\begin{theorem}\label{theo: increasing implies SIC}
Let $\cC$ be an increasing (resp. decreasing) flag code of type $(t_1, \dots, t_r)$ on $\bbF_q^n$. Then $\cC$ is a SIC flag code.
\end{theorem}
\begin{proof}
Let $\cC$ be an increasing flag code. Let us see that $\cC$ is a SIC flag code. To do so, assume that there exist flags $\cF, \cF'\in\cC$ with $\cF_{i-1} \subset \cF_i'$ for some $1< i \leq r$. Since $\cC$ is increasing, it is clear that $\cF_{i-1} =\cF'_{i-1}$. The same argument leads to $\cF_j=\cF'_j$ for every $1\leq j\leq i-1$. As a result, the flag
$$
(\cF_1, \dots, \cF_{i-1}, \cF'_i, \dots, \cF'_r)= \cF'\in\cC
$$
and then $\cC$ is SIC.

The proof for decreasing flag codes is similar.
\end{proof}

Combining Proposition \ref{prop: multiplicity increasing}, Theorem \ref{theo: increasing implies SIC} and Corollary \ref{cor:Cdeter_by_proj}, we come up with the next result.

\begin{theorem}\label{theo:increasing-deter}
If $\cC$ is an increasing (resp. decreasing) flag code, then it is determined by its projected codes.
\end{theorem}

The converse of Theorem \ref{theo:increasing-deter} is not true. It suffices to consider the flag code $\cC''$ in Example \ref{ex:Cnotdeterm}, which is determined by its projected codes but is neither increasing nor decreasing. The previous Theorem, together with Corollary 
\ref{cor:deter}, leads to the next result.
\begin{corollary}\label{cor: equivalence increasing decreasing}
    Let $\cC$ and $\cC'$ be flag codes of type $(t_1, \dots, t_r)$ on $\bbF_q^n$ and take $A\in \GL_n(\bbF_q)$ (resp. $(A,\varphi)\in \GaL_n(\bbF_q)$). If $\cC$ is increasing (or decreasing), then they are equivalent:
    \begin{enumerate}[$(i)$]
        \item $\cC'=\cC\cdot A$ (resp. $\cC'=\cC\cdot (A, \varphi)$), 
        \item $\cC'_i=\cC_i\cdot A$ (resp. $\cC'_i=\cC_i\cdot (A, \varphi)$), for every $1\leq i\leq r.$ 
    \end{enumerate}
\end{corollary}

 Notice that, by Theorem \ref{theo:increasing-deter}, every increasing (resp. decreasing) flag code is the unique flag code generated by its projected codes.  The rest of the section is devoted to exhibit families of flag codes that are increasing and/or decreasing and, consequently, determined by their sets of projected codes. More precisely, we will see that flag codes having projected codes of maximum distance are, in many cases, increasing and/or decreasing flag codes. In particular, we will derive some results concerning optimum distance flag codes. In our study, dimensions $t_a$ and $t_b$ in the type vector $(t_1,\dots, t_r)$ defined in (\ref{eq: dim ta and tb}), will play an important role. We start with the cases where there is only $t_a$ or $t_b$, that is, cases in which every dimension in the type vector of the flag code is, at least (resp. at most), $\frac{n}{2}$.

\begin{theorem}\label{th: max dist incresing decreasing} 
Let $\cC$ be a flag code of type $(t_1, \dots, t_r)$ on $\bbF_q^n$ having all its projected codes of maximum distance.
The following statements hold:
\begin{enumerate}[$(i)$]
\item If $t_1 \geq \frac{n}{2},$ then $\cC$ is increasing.
\item If $t_r \leq \frac{n}{2},$ then $\cC$ is decreasing.
\end{enumerate}
\end{theorem}
\begin{proof}

Assume that $t_1 \geq \frac{n}{2},$ take $1< i \leq r$ and a subspace $\cV\in\cC_{i}$ and suppose that $\cU, \cU'\in\cC_{i-1}$ satisfy $\cU\subset\cV$ and $\cU'\subset\cV$. In such a case, we have $\cU + \cU'\subset \cV \subsetneq \bbF_q^n$. Since $\cU$ and $\cU'$ belong to the subspace code $\cC_{i-1}$, having dimension  $t_{i-1}\geq \frac{n}{2}$ and maximum distance, by means of Lemma \ref{lem: max dist}, we conclude that $\cU=\cU'$, which proves that $\cC$ is  increasing.

On the other hand, if $t_r \leq \frac{n}{2}$, we consider an index  $1< i \leq r$ and a subspace $\cU\in\cC_{i-1}$. Let us prove that $\cU$ is contained in a single subspace in $\cC_i$. To do so, we consider subspaces $\cV, \cV'\in\cC_i$ and suppose that $\cU\subset\cV$  and  $\cU\subset\cV'$ hold. Thus, we have $\cU\subset \cV\cap\cV'.$ Besides, since $\cV, \cV'$  are elements in the constant dimension code $\cC_i$ of dimension 
$t_i\leq \frac{n}{2}$ and maximum distance, by Lemma \ref{lem: max dist}, we get $\cV=\cV'$. As a consequence, $\cC$ is a decreasing flag code.

\end{proof}

To address the remaining case, in which $t_1\leq \frac{n}{2} \leq t_r$, we will have to consider the two dimensions $t_a$ and $t_b$. Moreover, in this case, some extra conditions will be necessary in order to obtain the increasing or decreasing property.

\begin{theorem}\label{th: projected max dist increasing decreasing}
Let $\cC$ be a flag code of type $(t_1, \dots, t_r)$ on $\bbF_q^n$  such that $t_1\leq \frac{n}{2} \leq t_r$ and satisfying the following three conditions:
\begin{enumerate}[$(i)$]
\item All projected codes of $\cC$ are of maximum distance. 
\item $t_b < 2t_a$ (resp. $2t_b< n+t_a$) and 
\item $|\cC_1|=\dots=|\cC_a|$ (resp. $|\cC_b|=\dots=|\cC_r|$).
\end{enumerate}
Then $\cC$ is an increasing (resp. decreasing) flag code. 
\end{theorem}
\begin{proof}
Assume that $\cC_1, \dots ,\cC_r$ are constant dimension codes of maximum distance. Moreover, suppose that $t_b < 2t_a$ and  that $|\cC_1|=\dots=|\cC_a|$ hold. Let us see that $\cC$ is an increasing flag code. To this purpose,  for every $1< i\leq r$, we want to prove that, for every subspace $\cV\in \cC_i$, there exists a unique subspace $\cU\in\cC_{i-1}$ such that $\cU\subset\cV$. Equivalently, we show that it is possible to define the map $\alpha_i$ as in  (\ref{eq: map increasing}) such that  $\alpha_i(\cV)=\cU$. Assume that $\cU, \cU'\in\cC_{i-1}$ satisfy $\cU\subset\cV$  and $\cU'\subset \cV$. Let us see that $\cU=\cU'$.  We distinguish three cases for the index $i$:
\begin{enumerate}

\item If $i> b$, then we have $t_{i-1}\geq \frac{n}{2}$. Since $\cU + \cU'\subset \cV\subsetneq \bbF_q^n$ and $d_S(\cC_{i-1})$ is maximum, by Lemma \ref{lem: max dist} we conclude $\cU=\cU'$.

\item If $i=b>a$, then  $i-1=a$ and $t_{i-1}\leq \frac{n}{2}$. Since $\cC_{i-1}$ has maximum distance, Lemma \ref{lem: max dist} ensures that 
$$
\dim(\cU+\cU')= \left\lbrace
\begin{array}{lll}
t_{i-1}=t_a & \text{if} & \cU=\cU',\\
2t_{i-1}=2t_a & \text{if} & \cU\neq\cU'.
\end{array}
\right.
$$
On the other hand, $\dim(\cU+\cU')\leq \dim(\cV)= t_b < 2t_a$. As a consequence, $\cU=\cU'$ as stated. 

 \item If $i=b=a$ or $i<b$, then it follows that $i\leq a$ and $|\cC_i|=|\cC_{i-1}|$. Therefore, Remark \ref{rem:alpha} ensures us that the map $\alpha_i$ will exist if, and only if, the map  $\beta_i$ defined in (\ref{eq: map decreasing}) exists. Next we prove the existence of the map $\beta_i$. For this, we consider an arbitrary $\cX\in\cC_{i-1}$ and assume that there exist $\cY,\cY'\in\cC_{i}$ such that  $\cX\subseteq \cY$ and $\cX\subseteq \cY'$. Then $\dim(\cY\cap \cY')\geq \dim(\cX) = t_{i-1} > 0.$ However, since $i\leq a$ it follows that $t_i \leq  \frac{n}{2}$ and then $\cY=\cY'$ because $\cC_i$ is of maximum distance (see Lemma \ref{lem: max dist} for dimensions up to $\frac{n}{2}$). Thus, the map $\beta_i$ exists and its inverse map is just $\alpha_i$. In particular, the image of $\cV$ through $\alpha_i$ must be unique, that is, $\cU=\alpha_i(\cV)=\cU'$. 

\end{enumerate} 

Similar arguments can be used to prove the decreasing case.

\end{proof}

The conditions on the dimensions $t_a$ and $t_b$ in the previous result are needed, as we can see in the next example.

\begin{example}\label{ex: both conditions ta tb are needed}
For $n=10$, we consider the flag code $\cC$ of type $(4,7)$ consisting of the three following flags
$$
\begin{array}{rcl}
\cF^1 & = & (\left\langle e_1, e_2, e_3, e_4 \right\rangle, \left\langle e_1, e_2, e_3, e_4, e_5, e_6, e_7 \right\rangle), \\
\cF^2 & = & (\left\langle e_1, e_2, e_3, e_4 \right\rangle, \left\langle e_1, e_2, e_3, e_4, e_8, e_9, e_{10} \right\rangle), \\
\cF^3 & = & (\left\langle e_5, e_6, e_7, e_8 \right\rangle, \left\langle e_1, e_5, e_6, e_7, e_8, e_9, e_{10} \right\rangle). \\
\end{array}
$$ 
We have $t_1=t_a=4$ and $t_2=t_b=7$ and both $\cC_1$ and $\cC_2$ have maximum distance. Since $t_b = 7 < 2t_a=8,$ by means of Theorem \ref{th: projected max dist increasing decreasing}, the code is increasing. On the other hand, the condition $2t_b < n+t_a$ does not hold ($2t_b=14$ and $n+t_a=14$) and the code is not decreasing, since the subspace $\cF_1^1$ is contained in two different subspaces in $\cC_2$.
\end{example}

\begin{remark}
Notice that Theorems \ref{th: max dist incresing decreasing} and \ref{th: projected max dist increasing decreasing} can be applied for the particular case of optimum distance flag codes, showing that these kind of flag codes are increasing or decreasing under certain conditions on the type vector. Nevertheless, since optimum distance flag codes are in particular disjoint flag codes, we can obtain a stronger result by virtue of Theorem \ref{th: disjoint inc iff dec}. 
\end{remark}

The proof of the following result comes directly by applying Theorems \ref{th: max dist incresing decreasing} and \ref{th: projected max dist increasing decreasing}, together with Theorems \ref{th: disjoint inc iff dec} and \ref{theo:increasing-deter}.

\begin{corollary}\label{cor: odfc increasing and decreasing 2}
Consider an optimum distance flag code $\cC$ of type $(t_1, \dots, t_r)$ on $\bbF_q^n$ and assume that, at least, one of the following statements holds:
\begin{enumerate}[$(i)$]
\item $t_1 \geq \frac{n}{2}$,
\item $t_r\leq \frac{n}{2}$,
\item $t_1\leq \frac{n}{2}\leq t_r$ and $t_b < 2t_a$,
\item $t_1\leq \frac{n}{2}\leq t_r$ and $2t_b < n+t_a$;
\end{enumerate}
then $\cC$ is increasing and decreasing. %In any case, $\cC$ is determined by its projected codes.
\end{corollary}

Observe that, at least one of the conditions $t_b < 2t_a$ or $2t_b< n+t_a$ must be fulfilled if we want an optimum distance flag code to be increasing and decreasing, and then determined by its projected codes. It suffices to consider the flag code $\cC$ in Example \ref{ex: disjoint but not inc dec}, which is an optimum distance flag code but it is neither increasing nor decreasing and it is clearly not determined by its projected codes.

We finish the section with some results concerning (semi)linear equivalence for optimum distance flag codes. The proof of the next result comes  from Corollary 
 \ref{cor: equivalence increasing decreasing},  combined with  Corollary \ref{cor: odfc increasing and decreasing 2}.

\begin{corollary}
    Let $\cC$ be an optimum distance flag code of type $(t_1, \dots, t_r)$ on $\bbF_q^n$, with projected codes $\cC_1,\dots, \cC_r,$ and consider an element $A\in\GL_n(\bbF_q)$ (resp. $(A, \varphi)\in\GaL_n(\bbF_q)$). If $\cC'$ has projected codes $\cC_i\cdot A$ (resp. $\cC_i\cdot (A, \varphi))$ for every $1\leq i\leq r$ and, at least, one of the following statements holds:
\begin{enumerate}[$(i)$]
\item $t_1 \geq \frac{n}{2},$
\item $t_r\leq \frac{n}{2},$
\item $t_1\leq \frac{n}{2}\leq t_r$ and $t_b < 2t_a$,
\item $t_1\leq \frac{n}{2}\leq t_r$ and $2t_b < n+t_a$;
\end{enumerate}
then $\cC'=\cC\cdot A$ (resp. $\cC'=\cC\cdot (A,\varphi)$).
\end{corollary}

Notice that, at least one of the conditions in the previous list is required. Otherwise, as we can see in the next example, the result does not hold.
\begin{example}
Let $\{e_1, \dots, e_6\}$ be the standard $\bbF_q$-basis of $\bbF_q^6$ and consider the optimum distance flag code $\cC=\{\cF, \cF'\}$ of type $(2,4)$ on $\bbF_q^6$ given in Example \ref{ex: disjoint but not inc dec}, where $\cF$ and $\cF'$ are
$$
\begin{array}{ccc}
\cF    &=& (\langle e_1, e_2 \rangle, \langle e_1, e_2, e_3, e_4 \rangle)\\
\cF' &=& (\langle e_3, e_4 \rangle, \langle e_3, e_4, e_5, e_6 \rangle).
\end{array}
$$
Now consider a third flag $\cF''= (\langle e_3, e_4 \rangle, \langle e_1, e_2, e_3, e_4 \rangle)$ and form the flag code $\cC'=\{\cF, \cF', \cF''\}$. It is clear that $\cC'_i=\cC_i$, for $1\leq i\leq 2$. Nevertheless, $\cC$ and $\cC'$ cannot be equivalent since they have different cardinality (and also different minimum distance).
\end{example}

Concerning the automorphism group of optimum distance flag codes, the following result follows from  Corollary \ref{cor: odfc increasing and decreasing 2}, together with Theorems \ref{theo:igualaut} and \ref{theo: increasing implies SIC}.

\begin{corollary}
    Let $\cC$ be an optimum distance flag code of type $(t_1, \dots, t_r)$ on $\bbF_q^n$ and assume that, at least, one of the following statements holds:
\begin{enumerate}[$(i)$]
\item $t_1 \geq \frac{n}{2},$
\item $t_r\leq \frac{n}{2},$
\item $t_1\leq \frac{n}{2}\leq t_r$ and $t_b < 2t_a$,
\item $t_1\leq \frac{n}{2}\leq t_r$ and $2t_b < n+t_a$.
\end{enumerate}
Then $\Aut(\cC)=\bigcap_{i=1}^r \Aut(\cC_i)$ and $\SAut(\cC)=\bigcap_{i=1}^r \SAut(\cC_i).$
\end{corollary}

We end the section with an example showing that an optimum distance flag code can satisfy $\Aut(\cC)\neq\bigcap_{i=1}^r \Aut(\cC_i)$ if its type vector does not verify any of the conditions required by the previous corollary.

\begin{example}
Let $\{e_1, e_2, e_3, e_4\}$ denote the standard $\bbF_q$-basis of $\bbF_q^4$ and consider the type vector $(t_1, t_2)=(1,3)$ on $\bbF_q^4.$ Notice that for this type vector, we have $t_a=1$ and $t_b=3$ and conditions $t_b<2t_a$ and $2t_b<n+t_a$ do not hold. For this choice of the parameters, we form the optimum distance flag code $\cC=\{\cF, \cF'\}$, with flags
$$
\begin{array}{ccc}
    \cF  & = & (\langle e_1\rangle, \langle e_1, e_2, e_3\rangle ),  \\
    \cF' & = & (\langle e_2\rangle, \langle e_1, e_2, e_4\rangle ). 
\end{array}
$$
The matrix
$$
A=\begin{pmatrix}
    1 & 0 & 0 & 0\\
    0 & 1 & 0 & 0\\
    0 & 0 & 0 & 1\\
    0 & 0 & 1 & 0
\end{pmatrix}
$$
satisfies:
$$
\begin{array}{ccccccc}
\langle e_1\rangle\cdot A & = &  \langle e_1\rangle,  &  &  \langle e_1, e_2, e_3 \rangle\cdot A & = &  \langle e_1, e_2, e_4 \rangle,\\
\langle e_2\rangle\cdot A & = & \langle e_2\rangle, &   &\langle e_1, e_2, e_4 \rangle\cdot A & = &  \langle e_1, e_2, e_3 \rangle,
\end{array}
$$
and then, it is clear that $A\in\Aut(\cC_1)\cap\Aut(\cC_2).$ However, we have $A\notin\Aut(\cC)$, since $\cF\cdot A= (\langle e_1\rangle, \langle e_1, e_2, e_4\rangle )\notin\cC.$

\end{example}

\section{Conclusions}

In this paper we have introduced the notions of (semi)linearly equivalent flag codes. In particular, we have observed that  (semi)linearly equivalent flag codes have always (semi)linearly equivalent projected codes but the converse is not true in general. In this work, we have determined certain families of flag codes for which this converse holds true. To this end, we have developed a new approach to flag codes, starting from families of constant dimension codes and analyzing when they are suitable to provide (one or more) flag codes. 

Throughout this paper, we have defined new properties for flag codes that lead to the new families of SIC flag codes, flag codes determined by their projected codes and increasing/decreasing flag codes. These new families allow us to give examples of flag codes for which the (semi)linear equivalence of the projected codes guarantees the (semi)linear equivalence of the flag codes. Our study also provides results concerning the (semi)linear automorphism group of certain families of flag codes, including optimum distance flag codes.

\end{document}